\documentclass[aps,prb,twocolumn,amsmath,amssymb,eqsecnum]{revtex4-1}

\usepackage{amsmath}
\usepackage{amssymb}
\usepackage[pdftex]{color}
\usepackage{graphicx}
\usepackage{dcolumn} 
\usepackage{bm} 
\usepackage{longtable}
\usepackage{ulem}

\usepackage{enumitem}

\usepackage{color}
\usepackage{graphicx}
\usepackage{curves}
\usepackage{epic}
\usepackage{wasysym}
\usepackage{epsf, times}
\usepackage{subfigure}
\usepackage{eufrak}
\usepackage{amsthm}

\newcommand{\bs}[1]{{\boldsymbol{#1}}}

\newcommand{\ket}[1]{{|\,{#1}\,\rangle}}
\newcommand{\bra}[1]{{\langle\,{#1}\,|}}

\newcommand{\s}{{\sigma}}

\begin{document}

\author{Luiz H. Santos}
\affiliation
{
Department of Physics and Institute for Condensed Matter Theory,
University of Illinois at Urbana-Champaign, 1110 West Green Street, Urbana, Illinois, 61801-3080, USA
}

\author{Eduardo Fradkin}
\affiliation
{
Department of Physics and Institute for Condensed Matter Theory,
University of Illinois at Urbana-Champaign, 1110 West Green Street, Urbana, Illinois, 61801-3080, USA
}

\title{
     Instanton effects in lattice models of bosonic symmetry-protected topological states
      }

\begin{abstract}

Bosonic symmetry-protected topological (SPT) states are gapped disordered
phases of matter possessing symmetry-preserving boundary excitations.
It has been proposed that, at long wavelengths, the universal properties 
of an SPT system are captured by an effective non-linear sigma model field theory
in the presence of a quantized topological $\theta$-term.
%
By studying lattice models of bosonic SPT states, we are able to identify,
in their Euclidean path integral formulation,
(discrete) Berry phases that hold relevant physical information on the nature of the SPT ground states.
These discrete Berry phases are given intuitive physical interpretation  
in terms of instanton effects that capture the presence of a $\theta$-term
on the microscopic scale. 
\end{abstract}

\date{\today}

\maketitle



\section{Introduction}
\label{sec: introduction}

Since the prediction and discovery of topological 
band insulators,~\cite{Hasan-2010,Moore-2010,Qi-2011}
the relation between topology and symmetries in the realization of new
phases of matter has been the focus of intense scrutiny in recent years.
While the classification of non-interacting symmetry-protected topological (SPT)
fermionic phases of matter appears to have been completely formulated,~\cite{Schnyder-2008,Qi-2008,Kitaev-2009,Ryu-2010}
the quest for \textit{interacting} SPT phases is actively being theoretically 
studied.~\cite{Pollmann-2010,Schuch-2011,Chen-2011,Chen-2013,Ryu-2012,Levin-2012,Lu-2012,Vishwanath-2013,Xu-2013,Metliski-2013-a,Sule-2013,C-Wang-2014,Chen-2014,Santos-2014,Burnell-2013,J-Wang-2015,Bi-2013,Kapustin-2014,Santos-2015-a,Burnell-2015}
Moreover, it has been recently proposed that bosonic SPT states could 
be realized in periodically driven interacting systems,~\cite{Iadecola-2015}
as well as in other cold-atom platforms,~\cite{Liu-2014,Zhao-2015}
thus opening the interesting possibility to probe and manipulate SPT systems.

The simplest example of a bosonic SPT state is provided by the $S=1$
antiferromagnetic Heisenberg chain, whose ground state is gapped, symmetry
unbroken and possesses $2$-fold degenerate edge states that behave as $S=1/2$
low energy excitations.
Haldane has shown that the Euclidean path integral of the $S=1$ antiferromagnetic chain
is described by an Euclidean action that contains, in addition to a standard non-linear sigma model term, a topological $\theta$-term action $S_{\theta}$ with the coefficient $\theta$ quantized in multiples of $2 \pi$.~\cite{Haldane-1983-a,Haldane-1983-b}
\begin{equation}
\label{eq: O(3) NLSM def}
\begin{split}
S_{D = 1}
&\,=
S_{\rm{NLSM}}
+
\mathrm{i}
S_{\theta}
\\
&\,=
\int dx \, d\tau 
\Big[
\frac{1}{g}
(\partial_{\mu}\hat{n})^2
+
\mathrm{i}\,\frac{\theta}{8\pi}
\epsilon_{a b c} \epsilon_{\mu \nu} 
\hat{n}^{a} \partial_{\mu} \hat{n}^{b}  \partial_{\nu} \hat{n}^{c}
\Big]
\,.
\end{split}•
\end{equation}•
($\hat{n}$ is a $3$-component unit vector.)
Whereas the presence of the topological action does not change the 
partition function when periodic boundary conditions are imposed on the 
system 
[due to the fact that 
$
\exp{\left(\mathrm{i}\,S_{\theta}\right)} 
= 
\exp{\left(\mathrm{i}\,2\pi\times\rm{integer}\right)} 
= 
1
$], 
$S_{\theta}$ is nevertheless directly responsible for the $S=1/2$ excitation in the presence of edges.~\cite{Affleck-1987,Affleck-1988,Kennedy-1990,Hagiwara-1990,Fradkin-1992,Ng-1994}

Recently, Bi et al.~\cite{Bi-2013} proposed a classification of bosonic SPT states
in $D$-dimensional space via an extension of 
Eq.~(\ref{eq: O(3) NLSM def}),
whereby the gapped symmetric state is assumed to be described by
an O($D+2$) non-linear sigma model augmented with a quantized $\theta$-term action,
\begin{equation}
\label{eq: D dimensional field theory action}
\begin{split}
S_{D}
&\,=
S_{\rm{NLSM}}
+
\mathrm{i}
S_{\theta}
\\
&\,=
\int\,d^{D}x \, d\tau 
\Big[
\frac{1}{g}
(\partial_{\mu}\hat{n})^2
\\
&\,
+
\mathrm{i}\frac{2 \pi}{\Omega_{D+1}}
\epsilon_{a_{1}...a_{D+2}}
\hat{n}^{a_{1}}
\,
\partial_{x_{1}}\hat{n}^{a_{2}}
\,
...
\,
\partial_{x_{D}}\hat{n}^{a_{D+1}}
\,
\partial_{\tau}\hat{n}^{a_{D+2}}
\Big]
\,,
\end{split}•
\end{equation}•
where
$\Omega_{D+1}$ 
is the area of the $(D+1)$-dimensional sphere of unit radius.
In the strong coupling limit $g \rightarrow \infty$, the wave function acquires
the form~\cite{Xu-2013}
\begin{equation}
\label{eq: wave function NLSM}
\begin{split}
&\,
\ket{\Psi}
\sim
\int\,D\hat{n}(x)\,
e^
{
\mathrm{i}\frac{2 \pi}{\Omega_{D+1}}
\int d^{D}x\, \int^{1}_{0} du\,
\mathcal{W}[\hat{n}]
}
\,
\ket{\hat{n}(x)}
\\
&\,
\mathcal{W}[\hat{n}]
=
\epsilon_{a_{1}...a_{D+2}}
\hat{n}^{a_{1}}
\,
\partial_{x_{1}}\hat{n}^{a_{2}}
...
\,
\partial_{x_{D}}\hat{n}^{a_{D+1}}
\,
\partial_{u}\hat{n}^{a_{D+2}}
\,,
\end{split}•
\end{equation}•
where $\hat{n}(\bs{x},u)$ is an extension that satisfies
$\hat{n}(\bs{x},0) = (0,0,....,0,1)$
and
$\hat{n}(\bs{x},1) = \hat{n}(\bs{x})$.
The $\theta$-term action then endows the 
wave function with an amplitude given by a Wess-Zumino-Witten term
at level-$1$.~\cite{Wess-1971,Witten-1984}
Although the field theory approach adopted in Ref.~\onlinecite{Bi-2013} gives a useful platform for discriminating various classes of bosonic SPT states, 
there remains the question of how the properties encoded by the long wavelength
description Eq.~(\ref{eq: D dimensional field theory action})
are manifested at the microscopic scale. 

In this paper we investigate the effects of the $\theta$-term at the microscopic level, by studying the Euclidean partition function of microscopic Hamiltonians of bosonic SPT states. 
According to the work of Chen et al.,~\cite{Chen-2013}
SPT phases can be characterized by their ``short-range entanglement", 
in that an SPT ground state can be connected to a trivial state by the action of a unitary transformation that preserves the relevant global symmetry.
Recently, one of us,~\cite{Santos-2015-a}
using ideas of entanglement spectrum, 
has constructed explicit unitary transformations that give rise to one-dimensional SPT chains with time-reversal and
$\mathbb{Z}_{n}\times\mathbb{Z}_{n}$ symmetries, as well as 
two-dimensional SPT paramagnets with $\mathbb{Z}_{2}\times\mathbb{Z}_{2}$ symmetry, which are a generalization of the $\mathbb{Z}_{2}$ paramagnet
introduced by Levin and Gu in Ref.~\onlinecite{Levin-2012}.
In this work, we shall then use the unitary mappings studied in Ref.~\onlinecite{Santos-2015-a}
to find an explicit form of the Euclidean partition function for those classes of SPT states.

Expanding on the framework formulated by Chen and coworkers,\cite{Chen-2013} here we will investigate the structure of the cocycles of 1D and 2D spin SPT phases using a path-integral approach based on the standard mapping between quantum and classical spin systems using transfer matrices.\cite{Schultz-1964,Fradkin-1978} A recent treatment of 3D SPT phases has used a similarly inspired approach.\cite{Burnell-2015} Instead of a triangulation of the Euclidean space-time, as used in Ref.[\onlinecite{Chen-2013}], we will work with simple stacked  lattices in the Euclidean direction and show how the resulting effective discrete Euclidean action embodies the Berry phases of these SPT states. 
Thus, on our route to computing the partition function for the microscopic SPT models
considered here, we will be able to identify discrete Berry phases that originate from quantum fluctuations of the degrees of freedom, 
upon the evolution of the SPT system in imaginary time.
(For other studies on the role of Berry phase in SPT systems, 
see Refs.[\onlinecite{Liu-2013,Wang-2015-b,Gu-2016}])
To these discrete Berry phases, which can be viewed as instanton effects, we will attach a simple physical interpretation that will make the physics of the 
long wavelength $\theta$-term topological action manifest at the microscopic scale
in connection with the entanglement properties of the SPT
ground state.~\cite{Chen-2013,Santos-2015-a}

More specifically, after introducing our general approach in Section \ref{sec:general-approach}, we derive explicit forms in Section \ref{sec:examples} for the 1D time-reversal invariant and $\mathbb{Z}_n \times \mathbb{Z}_n$ invariant SPTs and for the 2D Levin-Gu model with $\mathbb{Z}_2$ symmetry and present the resulting Berry phases for each one of these cases.

\section{General approach}
\label{sec:general-approach}

We are interested in evaluating the partition function 
\begin{equation}
\label{eq: partition function}
\begin{split}
Z_{\rm{SPT}}
=
\mathrm{Tr}\,
\left(
e^
{
-\,\beta\,H_{\rm{SPT}}
}
\right)
\,,
\end{split}•
\end{equation}•
where the spin Hamiltonian $H_{\rm{SPT}}$ describes a bosonic SPT phase
in $D$-dimensional space. 
The partition function Eq.~(\ref{eq: partition function})
encodes $(D+1)$-dimensional space-time quantum fluctuations of the many-body system, with an Euclidean time direction of length $\beta = 1/T$ (inverse temperature) satisfying periodic
boundary conditions.

Our goal is to express the partition function Eq.~(\ref{eq: partition function})
in a local basis, in the process of which we will be able to identify non-trivial discrete
Berry phases originating from quantum fluctuations of the local spins.
These Berry phases will establish a simple and intuitive picture of the SPT state
from the point of view of the space-time quantum fluctuations of its microscopic 
degrees of freedom.

Important to our discussion is the fact that, if the $D$-dimensional
spatial manifold on which the SPT system lives has no boundaries,
the SPT Hamitonian $H_{\rm{SPT}}$ can be generated from a trivial gapped
paramagnetic Hamiltonian $H_{0}$ via a unitary, symmetry-preserving transformation $\mathbb{W}$:~\cite{Chen-2013,Santos-2015-a}
\begin{equation}
\label{eq: SPT Hamiltonian}
\begin{split}
H_{\rm{SPT}}
=
\mathbb{W}
\,
H_{0}
\,
\mathbb{W}^{-1}
\,,
\end{split}•
\end{equation}•
where $H_{0}$ describes a trivial paramagnet, i.e., a paramagnet whose
edge states can be gapped without symmetry violation.
In order to facilitate our obtaining of an explicit representation of the partition
function Eq.~(\ref{eq: partition function}), we choose to work with 
microscopic models in their zero correlation length limit 
[this choice will not affect the Berry phases, which are the subject of our attention
in Eq.~(\ref{eq: partition function})]:
\begin{equation}
\label{eq: trivial Hamiltonian}
\begin{split}
&\,
H_{0}
=
-
h\,
\sum^{N}_{i = 1}\,X_{i}
\,,
\\
&\,
[ X_{i} , X_{j} ] = 0,
\quad
\forall~ (i,j)
\,.
\end{split}•
\end{equation}•
$
N
$
is the number of lattice sites
and 
$
h
$
is a positive energy scale.
$X_{i}$ is a Hermitian operator defined solely on site $i$. 
(In its simplest form, $X_{i} = \sigma^{x}_{i}$ is a Pauli matrix.)
Due to the zero correlation form assumed for $H_{0}$, the Hamiltonian
Eq.~(\ref{eq: SPT Hamiltonian}) is a sum of mutually commuting operators,
\begin{equation}
\label{eq: SPT Hamiltonian general form}
\begin{split}
H_{\rm{SPT}}
=
-h\,\sum^{N}_{i = 1}\,
\mathcal{O}_{i}
=
-h\,\sum^{N}_{i = 1}\,
\mathbb{W}
\,
X_{i}
\,
\mathbb{W}^{-1}
\,.
\end{split}•
\end{equation}•
In order for the ground state of $H_{\rm{SPT}}$ to possess 
non-trivial entanglement, the transformation $\mathbb{W}$
can not be reduced to a product of on-site terms.
As a consequence, the operator $\mathcal{O}_{i}$ acts 
on site $i$ and the neighborhood thereof.

Despite the distinct entanglement patterns encoded by the 
ground states of $H_{0}$ and $H_{\rm{SPT}}$, when
expressed in the same local basis, the unitary transformation
Eq.~(\ref{eq: SPT Hamiltonian})
implies that in a closed spatial manifold, both Hamiltonians
have the same spectra and, hence, the same partition function.
Nevertheless, the fundamental physics of the SPT system can be
unveiled by studying the quantum fluctuations of spins between intermediate Euclidean
``time slices" of Eq.~(\ref{eq: partition function}).
These instanton events, as we will see by explicit computation, give rise to non-trivial phase factors in Eq.~(\ref{eq: partition function}), whereby spin fluctuations in imaginary time are coupled to domain wall like configurations in a way that is consistent with the underlying global symmetry of the SPT state.
We shall determine the phase factors associated to these instanton events 
for some cases of interest.

In order to carry out this program, we evaluate the trace in 
Eq.~(\ref{eq: partition function}) using the complete set of orthonormal 
many-body basis states $\ket{  \s  } = \ket{\s_1,\s_2,...,\s_{N}}$, whereby the 
trivial, unique ground state of $H_{0}$ is represented as
\begin{equation}
\label{eq: trivial ground state}
\begin{split}
\ket{\Psi_{0}}
=
\frac{1}{\mathcal{D}^{1/2}}
\,
\sum_{ \s } \, \ket{\s}
\,.
\end{split}•
\end{equation}•
$
\mathcal{D}
$
denotes the dimension of the Hilbert space and
$\ket{\Psi_{0}}$ is a product state
expressed in the ``ordered" basis $\ket{\s}$
satisfying
$
X_{j}\,\ket{\Psi_{0}} = \ket{\Psi_{0}}
$,
for every $j$.
In the ordered basis, all the diagonal matrix elements of $X_{j}$ vanish:
$
\bra{\s} X_{j} \ket{\s}
=
0
$.

The advantage of working with the representation Eq.~(\ref{eq: trivial ground state})
for the trivial ground state is that $\mathbb{W}$ is diagonal in the $\ket{\s}$ basis
:
$
\mathbb{W}
\,
\ket{\s}
=
e^
{
\mathrm{i}\,W(\s)
}
\,
\ket{\s}
$.
(See examples in Sec~\ref{sec:examples}.)
Hence, the SPT ground state in the zero correlation limit reads
\begin{equation}
\label{eq: SPT ground state}
\begin{split}
\ket{\Psi_{\rm{SPT}}}
=
\frac{1}{\mathcal{D}^{1/2}}
\,
\sum_{ \s } 
\, 
e^
{
\mathrm{i}\,W(\s)
}
\,
\ket{\s}
\,.
\end{split}•
\end{equation}•
Normalization factors aside,
$
e^
{
\mathrm{i}\,W(\s)
}
$
is the SPT ground state wavefunction in the $\ket{\s}$ basis.
Hence the phase factor $e^{\mathrm{i}W(\s)}$ in
Eq.~(\ref{eq: SPT ground state})
plays the role of the WZW term in 
Eq.~\ref{eq: wave function NLSM}.

\begin{figure}[h!]
\includegraphics[width=0.5\textwidth]{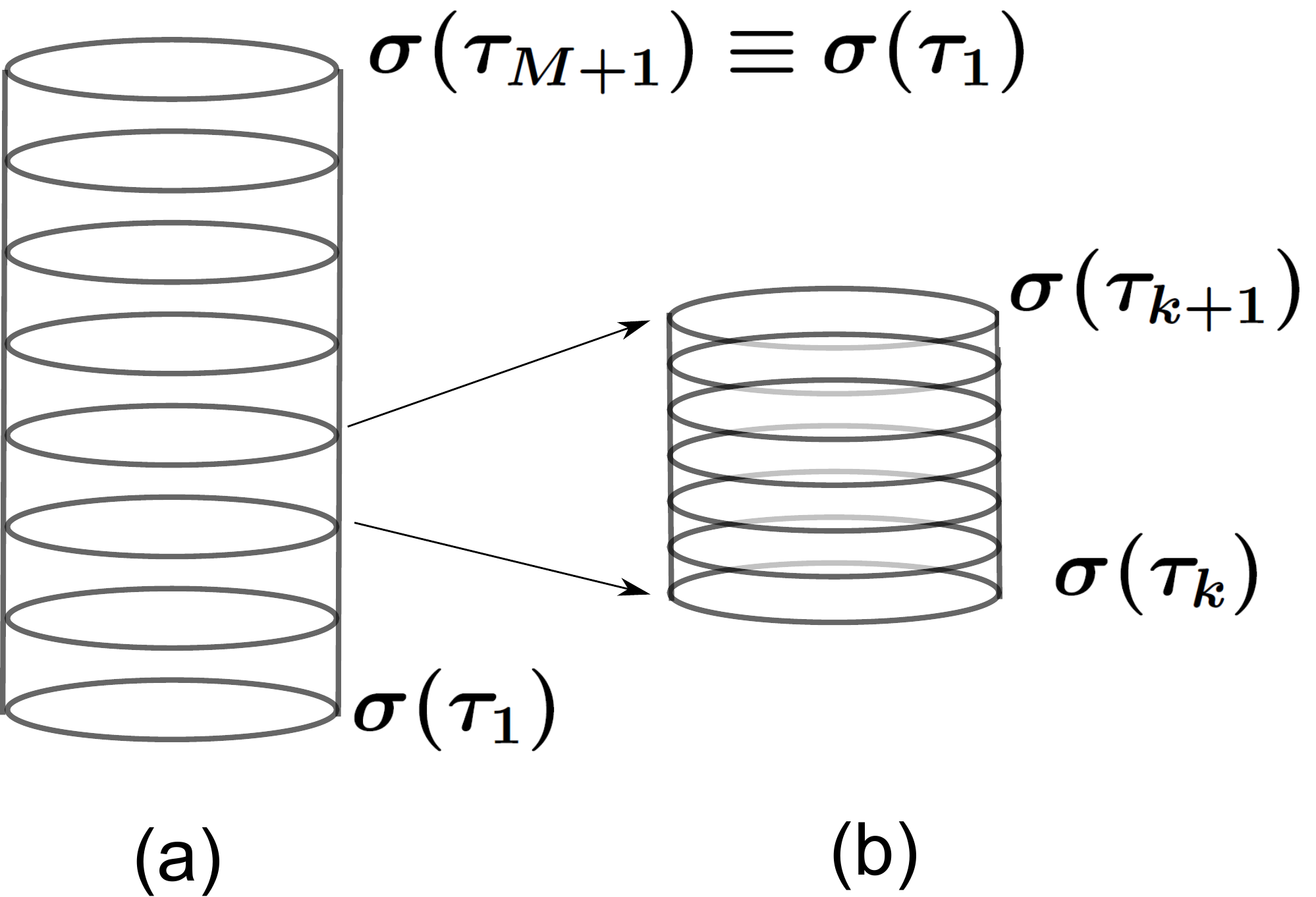}
\caption
{
Fig. 1(a) depicts the slicing of the partition function
into $M$ intervals, as described by
Eq.~(\ref{eq: time sliced partition function}).
At each time slice, we have an instantaneous representation
of the $D$-dimensional SPT system, which, without boundaries, is generically
represented by a circle.
Fig. 1(b) depicts the slicing of 
$
Z_{\rm{SPT}}
\Big[
\s(\tau_{k}),
\s(\tau_{k+1})
\Big]
$
into $N$ sub-intervals, as described by
Eq.~(\ref{eq: open surface partition function 2})
}
\label{fig:1}
\end{figure}

In order to explicitly capture the non-trivial quantum fluctuations
associated with the SPT Hamiltonian, we conventionally represent
the trace in  Eq.~(\ref{eq: partition function}) as:
\begin{subequations}
\begin{equation}
\label{eq: time sliced partition function}
\begin{split}
&\,
Z_{\rm{SPT}}
=
\sum_{\s(\tau_{1})}...\sum_{\s(\tau_{M})}
\prod^{M}_{k=1}
Z_{\rm{SPT}}
\Big[
\s(\tau_{k}),
\s(\tau_{k+1})
\Big]
\,,
\end{split}•
\end{equation}•
\begin{equation}
\label{eq: open surface partition function}
\begin{split}
&\,
Z_{\rm{SPT}}
\Big[
\s(\tau_{k}),
\s(\tau_{k+1})
\Big]
=
\bra{\s(\tau_{k})}
\,
e^
{
-\,\tau\,H_{\rm{SPT}}
}
\,
\ket{\s(\tau_{k+1})}
\\
&\,=
e^
{
\mathrm{i}\,
\Big[
W(\s(\tau_{k}))
-
W(\s(\tau_{k+1}))
\Big]
\,
}
\,
Z_{0}
\Big[
\s(\tau_{k}),
\s(\tau_{k+1})
\Big]
\\
&\,
\equiv
e^
{
\mathrm{i}\,
\Delta\,W_{k, k+1}
}
\,
Z_{0}
\Big[
\s(\tau_{k}),
\s(\tau_{k+1})
\Big]
\,,
\end{split}•
\end{equation}•
\end{subequations}•
where, in Eq.~(\ref{eq: time sliced partition function}), 
we have introduced $M$ time intervals of length 
$
\tau_{k+1} - \tau_{k}
=
\tau
=
\beta/M
$
with the periodic boundary condition in imaginary time
$
\s(\tau_{M+1}) \equiv \s(\tau_{1})
$
implied.
[See Fig.~\ref{fig:1}(a).]
Also, in Eq.~(\ref{eq: open surface partition function}) we have used the unitary transformation given in Eq.~(\ref{eq: SPT Hamiltonian}) to relate the imaginary time evolution of the SPT Hamiltonian at each time slice with
$
Z_{0}
\Big[
\s(\tau_{k}),
\s(\tau_{k+1})
\Big]
\equiv
\bra{\s(\tau_{k})}
\,
e^
{
-\,\tau\,H_{0}
}
\,
\ket{\s(\tau_{k+1})}
$.

Since
$
Z_{0}
\Big[
\s(\tau_{k}),
\s(\tau_{k+1})
\Big]
>
0
$,
all the non-trivial Berry phases associated with quantum fluctuations
of the SPT system are given by the phase factor on the second line
of Eq.~(\ref{eq: open surface partition function}).
Moreover, this phase factor has the interpretation of a surface term since it only depends on the configurations at the time slices $\tau_{k}$ and $\tau_{k+1}$.

While the phases appearing in
Eq.~(\ref{eq: open surface partition function})
account for the space-time quantum fluctuation of the whole system
(recall that $\s(\tau)$ refers to the many-body configuration at time $\tau$),
we can gain further information about the nature of the SPT system
by studying Berry phases picked up by \textit{local} spin fluctuations.
In order to do this, we divide each time interval $(\tau_{k}, \tau_{k+1})$ into
$N$ subintervals 
$
\left(
\tau_{k}, \tau_{k} + \epsilon, ...,  \tau_{k} +j \epsilon, ... , 
\tau_{k} + (N-1) \epsilon, \tau_{k+1}
\right)
\equiv
\left(
\tau_{k , 0}, \tau_{k,1}, ... , \tau_{k,j} , ... , \tau_{k, N-1}, \tau_{k, N}
\right)
$
of length $\epsilon = \tau/N = \beta/(M N)$ 
[see Fig~\ref{fig:1}(b)]
and rewrite
Eq.~(\ref{eq: open surface partition function}) as
\begin{equation}
\label{eq: open surface partition function version 2}
\begin{split}
&\,
Z_{\rm{SPT}}
\Big[
\s(\tau_{k}),
\s(\tau_{k+1})
\Big]
=
\sum_{\s(\tau_{k, 1})}
\sum_{\s(\tau_{k,2})}
...
\sum_{\s(\tau_{k, N-1})}
\\
&\,
\qquad
\times
\prod^{N}_{j = 1}\,
\bra{\s(\tau_{k , j-1})}
e^
{
h \epsilon \mathcal{O}_{j}
}
\ket{\s(\tau_{k , j})}
\,.
\end{split}•
\end{equation}•
In going from Eq.~(\ref{eq: open surface partition function})
to Eq.~(\ref{eq: open surface partition function version 2})
we have used the fact that the local operators $\mathcal{O}_{j}$ 
in Eq.~(\ref{eq: SPT Hamiltonian general form})
commute among themselves and we have introduced the identity operator $(N-1)$ times
in the form of a complete summation over intermediate many-body configurations
$
\s(\tau_{k,1}), ... , \s(\tau_{k, N-1})
$.

We are now faced with the evaluation of the transfer matrices
between many-body configurations of \textit{local} operators
$
e^
{
h \epsilon \mathcal{O}_{j}
}
$,
which, straightforwardly, yields
\begin{subequations}
\begin{equation}
\label{eq: transfer matrix of O operator}
\begin{split}
&\,
\bra{\s(\tau_{k,j-1})}
e^
{
h \epsilon \mathcal{O}_{j}
}
\ket{\s(\tau_{k, j})}
\\
&\,
=
e^
{
\mathrm{i}\,\mathcal{S}_{j}(\tau_{k,j-1} , \tau_{k}, j)
}
\,
\bra{\s_{j}(\tau_{k,j-1})}
e^
{
h \epsilon X_{j}
}
\ket{\s_{j}(\tau_{k, j})}
\,,
\end{split}•
\end{equation}•
\begin{equation}
\label{eq: single spin phase}
\begin{split}
e^
{
\mathrm{i}\,\mathcal{S}_{j}(\tau_{k,j-1} , \tau_{k}, j)
}
=
e^
{\mathrm{i}
\Big[
W(\s(\tau_{k,j-1}))
-
W(\s(\tau_{k,j}))
\Big]
}
\,
\Delta_{j}(\tau_{k,j-1},\tau_{k,j})
\,,
\end{split}•
\end{equation}•
\end{subequations}•
where
$
\Delta_{j}(\tau_{k,j-1},\tau_{k,j})
\equiv
\prod^{i \neq j}_{i}\,
\delta_{\s_{i}(\tau_{k,j-1}),\s_{i}(\tau_{k,j})}
$
enforces that all spins at time slices $\tau_{k, j-1}$ and $\tau_{k, j}$ 
be the same, except at site $j$.
Thus the phase
$
e^
{
\mathrm{i}\,\mathcal{S}_{j}(\tau_{k,j-1} , \tau_{k}, j)
}
$
in Eq.~(\ref{eq: single spin phase})
accounts for the Berry phase contribution 
due to the quantum fluctuations of a single spin at site $j$
in the presence of an instantaneous configuration of adjacent spins.
Therefore, the contribution of the partition function between time slices
$\tau_{k}$ and $\tau_{k+1}$,
which takes into account the Berry phases picked up by local spin
fluctuations, can be cast in the form
\begin{widetext}
\begin{equation}
\label{eq: open surface partition function 2}
\begin{split}
&\,
Z_{\rm{SPT}}
\Big[
\s(\tau_{k}),
\s(\tau_{k+1})
\Big]
=
\sum_{\s(\tau_{k, 1})}
\sum_{\s(\tau_{k,2})}
...
\sum_{\s(\tau_{k, N-1})}
\,
\prod^{N}_{j = 1}
\,
e^
{
\mathrm{i}\,
\mathcal{S}_{j}(\tau_{k,j-1} , \tau_{k}, j)
}
\bra{\s_{j}(\tau_{k,j-1})}
e^
{
h \epsilon X_{j}
}
\ket{\s_{j}(\tau_{k, j})}
\,.
\end{split}•
\end{equation}•
\end{widetext}

It is worthwhile to remind the reader that, even though,
by construction, the right hand sides
of Eq.~(\ref{eq: open surface partition function}) 
and Eq.~(\ref{eq: open surface partition function 2})
are identical,
the latter equation makes evident the Berry phases due to local instanton effects
while the former equation captures the quantum fluctuations of the entire $D$-dimensional system as it propagates in imaginary time.

\section{Examples}
\label{sec:examples}

\subsection{$D=1$, time-reversal symmetric SPT state}

A one-dimensional periodic SPT chain, invariant under
time-reversal $\mathbb{Z}^{T}_{2}$ operation,
\begin{equation}
\label{eq: time-reversal operator}
\Theta
=
\Big(
\prod^{N}_{j = 1}
\,
\s^{x}_{j}
\Big)\,K
\,,
\end{equation}•
($K$ denotes complex conjugation)
can be constructed using the unitary transformation~\cite{Santos-2015-a}
\begin{equation}
\label{eq: TRS unitary transformation}
\begin{split}
\mathbb{W}_{\rm{TRS}}
=
\prod^{N}_{j=1}
e^
{
\mathrm{i}\,\theta_{i,i+1}\,
\Big(
\frac
{
1
-
\s^{z}_{i}
\,
\s^{z}_{i+1}
}
{
2
}
\Big)
}
\,,
\quad
\theta_{i,i+1} = \frac{\pi}{2}
\,,
\end{split}•
\end{equation}•
where at every site of the chain there is a spin-$1/2$ 
degree of freedom represented by a Pauli operator 
$
\s^{a}_{i}
$
with
$
a = 1,2,3 = x, y, z
$.
The unitary operator Eq.~(\ref{eq: TRS unitary transformation})
endows a many-body basis state
$
\ket{\s}
=
\ket{\s_1,\s_2,...,\s_{N}}
$
with a phase factor 
$
\exp
{
\Big\{
\mathrm{i}
(\pi/2)
N_{\rm{d}}(\s)
\Big\}
}
=
\pm
1
$,
where
$N_{\rm{d}}(\s)$ denotes the (even) number of domain walls in the state $\ket{\s}$.
Moreover, this transformation commutes with the time-reversal operator
Eq.~(\ref{eq: time-reversal operator})
and each local unitary piece creates a maximally entangled
state between nearest neighbor spins.~\cite{Santos-2015-a}

So, under Eq.~(\ref{eq: time-reversal operator}), one can map 
the trivial time-reversal symmetric ground state
\begin{subequations}
\begin{equation}
\begin{split}
\ket{\Psi_{0}}
=
\Big(
\frac{\ket{\uparrow}+\ket{\downarrow}}{\sqrt{2}}
\Big)^{{\otimes}^{N}}
=
\frac{1}{2^{N/2}}
\,
\sum_{\s}\, \ket{\s}
\,,
\end{split}•
\end{equation}•
of
\begin{equation}
H_{0}
=
-h\,\sum^{N}_{j=1}\,\sigma^{x}_{j}
\,,
\end{equation}•
\end{subequations}•
into
\begin{subequations}
\begin{equation}
\begin{split}
\ket{\Psi_{\rm{TRS}}}
=
\mathbb{W}_{\rm{TRS}}\,\ket{\Psi_{0}}
=
\frac{1}{2^{N/2}}
\,
\sum_{\s}\, 
e^
{
\mathrm{i}
\frac
{
\pi
}
{
2
}
N_{\rm{d}}(\s)
}
\ket{\s}
\,,
\end{split}•
\end{equation}•
which is the unique ground state of the SPT Hamiltonian
\begin{equation}
\label{eq: time-reversal symmetric SPT Hamiltonian}
\begin{split}
H_{\rm{SPT}}
=
\mathbb{W}_{\rm{TRS}}
\,
H_{0}
\,
\mathbb{W}_{\rm{TRS}}^{-1}
=
h\,
\sum^{N}_{j = 1}\,
\s^{z}_{j-1}\,\s^{x}_{j}\,\s^{z}_{j+1}
\,.
\end{split}•
\end{equation}•
\end{subequations}•
It is immediate to see that the SPT Hamiltonian  
Eq.~(\ref{eq: time-reversal symmetric SPT Hamiltonian}), when open boundary conditions are imposed, possesses $2$-fold degenerate states per edge.

Now, applying our discussion of Sec. \ref{sec:general-approach}
to the time-reversal symmetric SPT Hamiltonian 
Eq.~(\ref{eq: time-reversal symmetric SPT Hamiltonian})
yields the phase contribution Eq.~(\ref{eq: single spin phase})
due to a single spin fluctuation to be
\begin{equation}
\label{eq: single spin phase TRS}
\begin{split}
&\,
e^
{
\mathrm{i}\,\mathcal{S}_{j}(\tau_{k,j-1} , \tau_{k}, j)
}
=
\\
&\,
e^
{
\mathrm{i}\,\pi
\Big(
\frac
{
\s_{j}(\tau_{k,j-1})
-
\s_{j}(\tau_{k,j})
}
{
2
}
\Big)
\,
\Big(
\frac
{
\s_{j+1}(\tau_{k,j})
-
\s_{j-1}(\tau_{k,j})
}
{
2
}
\Big)
}
\,,
\end{split}•
\end{equation}•
[the product of delta functions 
$
\Delta_{j}(\tau_{k,j-1},\tau_{k,j})
$
in Eq.~(\ref{eq: single spin phase})
is omitted in Eq.~(\ref{eq: single spin phase TRS})].
Combining the result Eq.~(\ref{eq: single spin phase TRS}) 
due to single spin processes yields, according to 
Eq.~(\ref{eq: open surface partition function}),
the total Berry phase as the one-dimensional SPT chain evolves in imaginary time:
\begin{equation}
\label{eq: open surface phase TRS}
\begin{split}
&\,
e^
{
\mathrm{i}\,
\Delta\,W_{k,k+1}
}
=
e^
{
\mathrm{i}\,\frac{\pi}{2}
\Big[
N_{\rm{d}}(\s(\tau_{k}))
-
N_{\rm{d}}(\s(\tau_{k+1}))
\Big]
\,
}
\,.
\end{split}•
\end{equation}•

Eq.~(\ref{eq: single spin phase TRS}) implies that a single spin
fluctuation at site $j$ contributes a phase $-1$ to the partition
function if the neighbor spins are antiparallel and $+1$ if they are
parallel to each other.
Summing up these individual phase contributions throughout the chain
gives
$
e^
{
\mathrm{i}\,
\Delta\,W_{k,k+1}
}
=
-1
$
if the change in domain wall number is $4m+2$ ($m \in \mathbb{Z}$) and 
$
e^
{
\mathrm{i}\,
\Delta\,W_{k,k+1}
}
=
+1
$
if the change in domain wall number is $4m$.

\subsection
{
$D=1$, $\mathbb{Z}_{n}\times\mathbb{Z}_{n}$ symmetric SPT state
}

We consider a periodic chain with an even number $N$ of sites where, at every site $j$,
there exists a clock operator $\s_{j}$ and its conjugate operator $\tau_{j}$ satisfying
the algebra
\begin{equation}
\label{eq: Zn algebra}
\begin{split}
&\,
\s^{n}_{j}
=
\tau^{n}_{j}
=
1
\,,
\quad
\s^{\dagger}_{j}
=
\s^{n-1}_{j}
\,,
\tau^{\dagger}_{j}
=
\tau^{n-1}_{j}
\,,
\\
&\,
\tau^{\dagger}_{j}\,
\s_{j}\,
\tau_{j}
=
\omega\,
\s_{j}
\,,
\quad
\omega
\equiv
e^
{
i\,\frac{2\,\pi}{n}
}
\,.
\end{split}•
\end{equation}•
For the $n=2$ case, Eq.~(\ref{eq: Zn algebra}) admits 
a Hermitian representation in terms of Pauli matrices  
$\s_{j} = \s^{z}_{j}$ and $\tau_{j} = \s^{x}_{j}$;
otherwise these clock operators are not Hermitian
(see Ref.~\onlinecite{Fendley-2012} for a recent discussion).
We shall denote by $\ket{\s_{i}}$,
for
$
\s_{i}
\in
\{ 1, \omega, ..., \omega^{n-1} \}
$,
the eigenstates of $\s_{i}$,
and by
$\ket{\s} = \ket{\s_{1}, ... ,\s_{N}}$
the corresponding many-body state.
(The distinction between operators and their eigenvalues should be clear from 
the context.)

We choose to work with a representation in which the generators of the $\mathbb{Z}_{n}\,\times\,\mathbb{Z}_{n}$ 
symmetry implement transformations
$
\s_{j} \rightarrow \omega\,\s_{j}
$
independently for the operators on even and odd sublattices,
and hence are given by
\begin{equation}
\label{eq: Zn times Zn symmetry in one dimension}
\begin{split}
\widehat{S}^{(1)}_
{
\mathbb{Z}_{n}
}
=
\prod_{j \in \mathrm{even}}\,
\tau_{j}
\,,
\quad
\widehat{S}^{(2)}_
{
\mathbb{Z}_{n}
}
=
\prod_{j \in \mathrm{odd}}\,
\tau_{j}
\,.
\end{split}•
\end{equation}•

A trivial ground state and its parent Hamiltonian, both invariant
under the action of the operators in Eq.~(\ref{eq: Zn times Zn symmetry in one dimension}),
are given by
\begin{subequations}
\begin{equation}
\label{eq: Trivial Zn times Zn ground state in one dimension}
\begin{split}
\ket{\Psi_{0}}
&\,=
\Big(
\frac
{
\ket{1}
+
\ket{\omega}
+
...
+
\ket{\omega^{n-1}}
}
{
\sqrt{n}
}
\Big)^{{\otimes}^{N}}
\\
&\,=
\frac
{
1
}
{
n^{N/2}
}
\,
\sum_{\{ \s \}}\,
\ket{\s}
\,,
\end{split}•
\end{equation}•
\begin{equation}
\label{eq: Trivial Hamiltonian for the Zn times Zn state in one dimension}
H_{0}
=
-h\,
\,
\sum^{N}_{j=1}\,
\left(
\tau_{j}
+
\tau^{\dagger}_{j}
\right)
\,.
\end{equation}•
\end{subequations}•

There exist $n-1$ unitary transformations 
$
\mathbb{W}_{n}^{(p)}
$,
$p \in \{ 1, ..., n-1 \}$,
that (i) map the product state 
Eq.~(\ref{eq: Trivial Zn times Zn ground state in one dimension}) 
into a new state where every spin is maximally entangled 
with its nearest neighbors and (ii) commutes with the symmetries 
Eq.~(\ref{eq: Zn times Zn symmetry in one dimension}):~\cite{Santos-2015-a}
\begin{equation}
\label{eq: Particular form of the unitary mapping Zn times Zn in one dimension}
\begin{split}
&\,
\mathbb{W}_{n}^{(p)}
=
e^
{
i\,\frac{2 \pi p}{n}
\,
\sum_{j}
\sum^{n-1}_{a=1}\,
\frac
{
(
\s^{\dagger}_{2j}\,
\s_{2j+1}
)^a
-
(
\s^{\dagger}_{2j-1}\,
\s_{2j}
)^a
}
{
(
\omega^a - 1
)
\,
(
\bar{\omega}^a - 1
)
}
}
\\
&\,
\mathbb{W}_{n}^{(p)}
\,
\ket{\s}
\equiv
e^
{
\mathrm{i}\,W^{(p)}_{n}(\s)
}
\,
\ket{\s}
\,,
\quad
W^{(p)}_{n}(\s)
\in
\mathbb{R}
\,.
\end{split}•
\end{equation}•
Each of the 
$
\mathbb{W}_{n}^{(p)}
$
for $p \neq 0$ gives rise to an SPT ground state, and hence,
there are $n-1$ SPT classes.~\cite{Chen-2013,Bi-2013}

With the transformation 
Eq.~(\ref{eq: Particular form of the unitary mapping Zn times Zn in one dimension}),
one then arrives at the SPT Hamiltonian~\cite{Santos-2015-a}

\begin{equation}
\label{eq: Non-trivial Zn times Zn Hamiltonian in one dimension}
\begin{split}
&\,
H^{(p)}_{n}
=
\mathbb{W}_{n}^{(p)}
\,
H_{0}
\,
\left(
\mathbb{W}_{n}^{(p)}
\right)^{-1}
\\
&\,=
-h\,
\,
\sum_{j}\,
\Big\{
\Big[
\tau_{2j}
\,
(
\s_{2j-1}\,
\s^{\dagger}_{2j+1}
)^{p}
+
\tau_{2j+1}
\,
(
\s^{\dagger}_{2j}\,
\s_{2j+2}
)^{p}
\Big]
\\
&\,
\qquad\qquad\qquad
+
\textrm{H.c.}
\Big\}
\end{split}
\end{equation}
%
and its ground state
\begin{equation}
\begin{split}
\ket{\Psi^{(p)}_{n}}
=
\mathbb{W}_{n}^{(p)}
\,
\ket{\Psi_{0}}
=
\frac
{
1
}
{
n^{N/2}
}
\,
\sum_{\{ \s \}}\,
e^
{
\mathrm{i}\,W^{(p)}_{n}(\s)
}
\ket{\s}
\,.
\end{split}•
\end{equation}•

The SPT Hamiltonian 
Eq.~(\ref{eq: Non-trivial Zn times Zn Hamiltonian in one dimension}), when open boundary conditions are imposed, possesses $n$-fold degenerate states per edge.

Now, applying our discussion of Sec. \ref{sec:general-approach}
to the SPT Hamiltonian 
Eq.~(\ref{eq: Non-trivial Zn times Zn Hamiltonian in one dimension})
yields the phase contribution Eq.~(\ref{eq: single spin phase})
due to a single spin fluctuation to be
\begin{equation}
\label{eq: single spin phase Zn times Zn}
\begin{split}
&\,
e^
{
\mathrm{i}\,\mathcal{S}_{j}(\tau_{k,j-1} , \tau_{k}, j)
}
=
\\
&\,
e^
{
\eta_{j}\,\mathrm{i}\,\frac{2 \pi}{n} p\,
\sum^{n-1}_{a = 1}
\Big[
\frac
{
\bar{\s}_{j}^{a}(\tau_{k,j-1})
-
\bar{\s}_{j}^{a}(\tau_{k,j})  
}
{
\bar{\omega}^{a} - 1
}
\Big]
\Big[
\frac
{
{\s^{a}_{j+1}}(\tau_{k,j})
-
{\s^{a}_{j-1}}(\tau_{k,j})  
}
{
\omega^{a} - 1
}
\Big]
}
\,,
\end{split}•
\end{equation}•
where
$
\eta_{j} = 1
$
if $j$ is even, and 
$
\eta_{j} = -1
$
if $j$ is odd.
[the product of delta functions 
$
\Delta_{j}(\tau_{k,j-1},\tau_{k,j})
$
in Eq.~(\ref{eq: single spin phase})
is omitted in Eq.~(\ref{eq: single spin phase Zn times Zn})].
According to Eq.~(\ref{eq: single spin phase Zn times Zn}) the Berry phases due
to a single spin fluctuation in imaginary are non-zero provided 
the neighbor spins are not parallel to each other. 
Using the expression 
Eq.~(\ref{eq: Identity Zn operators}) 
in the appendix, one can 
show that if the spin at site $j$ fluctuates between values 
$\s_{j}(\tau_{k,j-1}) = \omega^{\ell_{0}+\ell}$
and
$\s_{j}(\tau_{k,j}) = \omega^{\ell_{0}}$,
for $\ell_{0},\ell \in \{0, ..., n-1 \}$,
then the expression 
Eq.~(\ref{eq: single spin phase Zn times Zn}) 
reduces to the simple form
$
e^
{
\mathrm{i}\,\mathcal{S}_{j}(\tau_{k,j-1} , \tau_{k}, j)
}
=
(\bar{\s}_{j+1}\,\s_{j-1})^{\eta_{j}\,\ell\,p}
$.

Combining the result 
Eq.~(\ref{eq: single spin phase Zn times Zn}) 
due to single spin processes yields, according to 
Eq.~(\ref{eq: open surface partition function}),
the total Berry phase as the one-dimensional 
$\mathbb{Z}_{n}\times\mathbb{Z}_{n}$ symmetric
SPT system evolves in imaginary time:
\begin{equation}
\label{eq: open surface phase Zn x Zn}
\begin{split}
&\,
e^
{
\mathrm{i}\,
\Delta\,W_{k,k+1}
}
=
e^
{
\mathrm{i}\,
\Big[
W^{(p)}_{n}(\s(\tau_{k}))
-
W^{(p)}_{n}(\s(\tau_{k+1}))
\Big]
\,
}
\,.
\end{split}•
\end{equation}•

\subsection{$D=2$, $\mathbb{Z}_{2}$ symmetric SPT state}

We now analyze the $2$D $\mathbb{Z}_{2}$ SPT model proposed by
Levin and Gu.~\cite{Levin-2012} 
This is an exactly solvable model where spin-$1/2$ degrees
of freedom are defined on the vertices of a triangular lattice. The Hamiltonian
of the system is
\begin{equation}
\label{eq: Levin-Gu Hamiltonian}
\begin{split}
H_{\mathbb{Z}_{2}}
=
h\,
\sum_{j}\,
\s^{x}_{j}\,
e^
{
\mathrm{i} \frac{\pi}{4}
\sum^{j}_{< \ell, \ell' >}\,
\left(
1 - \s^{z}_{\ell} \s^{z}_{\ell'}
\right)
}
\,,
\quad
h >0 
\,,
\end{split}•
\end{equation}•
where the summation 
$
\sum^{j}_{< \ell, \ell' >}
$
extends over nearest neighbor spins around the site $j$.

The ground state of the Levin-Gu model is
\begin{equation}
\ket{\Psi_{\mathbb{Z}_{2}}}
=
\frac{1}{2^{N/2}}
\sum_{\s}\,
(-1)^{L(\s)}
\ket{\s}
\,,
\end{equation}•
where $L(\s)$ counts the number of loops, defined in the dual (hexagonal) lattice,
associated with domain wall configurations of the the many-body state $\ket{\s}$ in the $\s^{z}$ basis.

There exists a unitary transformation $\mathbb{W}_{\mathbb{Z}_{2}}$,
connecting the Levin-Gu Hamiltonian to a trivial $2D$ paramagnetic Hamiltonian
\begin{equation}
\label{eq: trivial 2d Hamiltonian}
H_{0}
=
-h\,
\sum_{j}\,\s^{x}_{j}
\,,
\end{equation}•
whose ground state is a simple product state
\begin{equation}
\ket{\Psi_{0}}
=
\frac{1}{2^{N/2}}
\sum_{\s}\,
\ket{\s}
\,.
\end{equation}•
Such transformation, having the property 
\begin{equation}
\mathbb{W}_{\mathbb{Z}_{2}}
\,
\ket{\s}
=
(-1)^{L(\s)}
\,
\ket{\s}
\,,
\end{equation}•
reads
\begin{equation}
\label{eq: Z2 transformation}
\begin{split}
\mathbb{W}_{\mathbb{Z}_{2}}
&\,
=
\prod_{j}
\,
e^
{
-\mathrm{i}\,\frac{\pi}{6}
\,
\s^{z}_{j}
\,
D_{j}(\{ \s^{z} \})
}
\,,
\end{split}•
\end{equation}•
where
$
D_{j}(\{ \s^{z} \})
=
\sum^{j}_{<\ell \ell'>}
\Big(
\frac
{
1
-
\s^{z}_{\ell}
\s^{z}_{\ell'}
}
{
2
}
\Big)
$
defines the domain wall operator around the close loop 
formed by the six sites nearest neighbors of site $j$.

One verifies that 
\begin{equation}
\label{eq: transformed spin operators}
\begin{split}
&\,
\pi^{x}_{j} 
\equiv 
\mathbb{W}_{\mathbb{Z}_{2}} \, \s^{x}_{j} \, \mathbb{W}_{\mathbb{Z}_{2}}^{-1}
=
-
\s^{x}_{j}\,
e^
{
\mathrm{i} \frac{\pi}{4}
\sum_{< i k >; j}\,
\left(
1 - \s^{z}_{i} \s^{z}_{k}
\right)
}
\,,
\\
&\,
\pi^{y}_{j} 
\equiv 
\mathbb{W}_{\mathbb{Z}_{2}}  \, \s^{y}_{j}  \, \mathbb{W}_{\mathbb{Z}_{2}}^{-1}
=
-
\s^{y}_{j}\,
e^
{
\mathrm{i} \frac{\pi}{4}
\sum_{< i k >; j}\,
\left(
1 - \s^{z}_{i} \s^{z}_{k}
\right)
}
\,,
\\
&\,
\pi^{z}_{j} 
\equiv 
\mathbb{W}_{\mathbb{Z}_{2}}  \, \s^{z}_{j} \, \mathbb{W}_{\mathbb{Z}_{2}}^{-1}
=
\s^{z}_{j}
\,.
\end{split}•
\end{equation}•

As seen in Eq.~(\ref{eq: transformed spin operators}), the unitary transformation Eq.~(\ref{eq: Z2 transformation}) gives rise to a new set of Pauli operators whose phase factors depend 
on the domain wall operator 
$
D_{j}(\{ \s^{z} \})
$
surrounding site $j$.
Notice that since this domain wall operator takes even integer values,
the phase factors in Eq.~(\ref{eq: transformed spin operators}),
and hence $\pi^{a}_{j}$, are Hermitian.

From the explicit form of the unitary transformation
Eq.~(\ref{eq: Z2 transformation}),
we find, according to 
Eq.~(\ref{eq: single spin phase}), 
that the Berry phase contribution due to a single spin
fluctuation at site $j$ is given by
\begin{equation}
\label{eq: single spin phase Z2}
\begin{split}
&\,
e^
{
\mathrm{i}\,\mathcal{S}_{j}(\tau_{k,j-1} , \tau_{k}, j)
}
=
\\
&\,
e^
{
-\mathrm{i} \pi
\left(
\frac{\s_{j}(\tau_{k,j-1})-\s_{j}(\tau_{k,j})}{2}
\right)
\,
\Big[
1
+
\frac{1}{2}
\sum^{j}_{<\ell, \ell'>}
\left(
\frac
{
1
-
\s_{\ell}(\tau_{k,j}) \s_{\ell'}(\tau_{k,j})
}
{
2
}
\right)
\Big]
}
\\
&\,=
e^
{
-\mathrm{i} \pi
\left(
\frac{\s_{j}(\tau_{k,j-1})-\s_{j}(\tau_{k,j})}{2}
\right)
\,
\Big(
1
+
\frac{1}{2}
D_{j}(\tau_{k,j})
\Big)
}
\,.
\end{split}•
\end{equation}•

\begin{figure}[h!]
\includegraphics[width=0.5\textwidth]{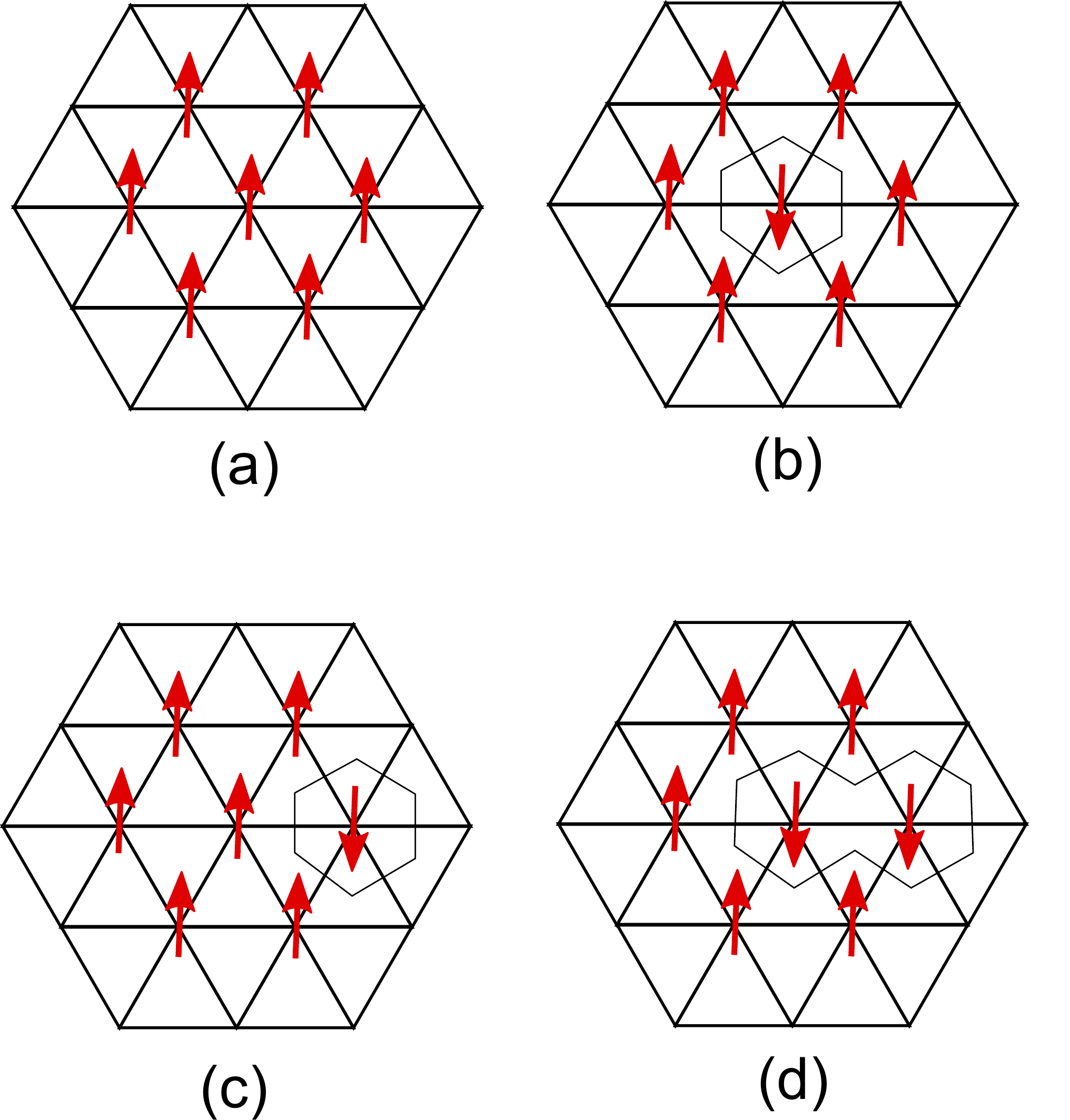}
\caption
{
Figures 2(a) and 2(b) capture the fluctuation of the middle spin
giving a phase $e^{\mathrm{i}\mathcal{S}_{j}} = -1$
[Eq.~(\ref{eq: single spin phase Z2})], 
as the configuration
of nearest neighbor spins has $D_{j} = 0$.
Figures 2(c) and 2(d) capture the fluctuation of the middle spin
giving a phase $e^{\mathrm{i}\mathcal{S}_{j}} = +1$
[Eq.~(\ref{eq: single spin phase Z2})], 
as the configuration
of nearest neighbor spins has $D_{j} = 2$.
}
\label{fig:2}
\end{figure}

Hence, the spin fluctuation at site $j$ contributes with $-1$ to the 
partition function if the configuration of surrounding spins has 
$
D_{j}
=
\{0,4\}
$,
while it contributes with $+1$ to the 
partition function if the configuration of surrounding spins has 
$
D_{j}
=
\{2, 6\}
$
(see Fig.~\ref{fig:2}).
Moreover, as the full two dimensional system evolves in imaginary 
time, it picks a phase factor
\begin{equation}
\label{eq: open surface phase Z2}
\begin{split}
&\,
e^
{
\mathrm{i}\,
\Delta\,W_{k,k+1}
}
=
e^
{
\mathrm{i}\,\pi
\Big[
L(\s(\tau_{k}))
-
L(\s(\tau_{k+1}))
\Big]
\,
}
\,,
\end{split}•
\end{equation}•
which is $-1$ if $\Delta\,L = 1$ (mod $2$), or
$+1$ if $\Delta\,L = 0$ (mod $2$).

\section{Summary and Discussion}
\label{sec: summary and discussion}

We have studied the path integral of bosonic SPT systems, focusing on the manifestation of the so called topological $\theta$-terms on the lattice scale.
We did so by investigating lattice models of bosonic SPT states,
which allowed us to compute the Berry phase contributions appearing in the
path integral due to the quantum fluctuations of local degrees of freedom.
In the examples we have considered, these non-trivial Berry phases
involve the coupling of local spin fluctuations with surrounding domain wall
like configurations, and thus illustrate, in a intuitive way, the connection between 
Berry phase effects and the non-trivial entanglement characteristic of SPT states.

Although here we have focused entirely on bosonic systems,
we close by mentioning some works that address the character
of Berry phases in fermionic systems.
At the level of relativistic free field theories (coupled to gauge fields) this problem has been studied in the high-energy literature in the context of anomalies and obstruction to gauging certain symmetries.~\cite{Niemi-1985,Sonoda-1986,Stone-1988}
It has been known from that work that there is indeed a connection between these anomalies and Berry phases. 
Also, a recent discussion of Berry phases in 
fermionic SPT systems that can be described by band theory and free fermions
has been given in Ref.~\onlinecite{Witten-2015}.
The bosonic cases studied here are strongly coupled and cannot be examined by the same methods as in the field-theoretic anomalies.
The nature of Berry phases in strongly coupled fermionic
systems is an interesting open problem.

\begin{acknowledgments}
This work was supported in part by the Gordon and Betty
Moore Foundation's EPiQS Initiative through Grant No.
GBMF4305 at the University of Illinois (LS) and by the National Science Foundation through grant No. DMR 1408713 at the University of Illinois (EF).
\end{acknowledgments}

\appendix
\section{$\mathbb{Z}_{n}$ operators}
\label{appendix}

A possible representation for the $(\s,\tau)$ operators
satisfying Eq.~(\ref{eq: Zn algebra}) is as follows
\begin{equation}
\s_{j} 
=
\begin{pmatrix} 
1 & 0 & 0 & 0 \\
0 & \omega & 0 & 0\\  
0 & 0 & \ddots  & 0\\  
0 & 0 & 0 & \omega^{n-1}
\end{pmatrix}
\,,
\quad
\tau_{j} 
=
\begin{pmatrix} 
0 & 0 & 0 & \dots &0& 1  \\
1 & 0 & 0 & \dots &0& 0 \\
0 & 1 & 0 & \dots &0& 0 \\
0 & 0 & 1 & \dots &0& 0 \\
\vdots &0 & 0 & \dots &1 & 0 
\end{pmatrix} 
\,,
\end{equation}•
where $\s$ is a clock variable (in the diagonal representation)
and $\tau$ is a raising and lowering operator.

Consider the following Hermitian operator:
\begin{equation}
\label{eq: definition of q}
\begin{split}
q(\s)
=
\frac{n-1}{2}
+
\sum^{n-1}_{a=1}\,
\frac
{
\s^a
}
{
\bar{\omega}^a - 1
}
\,.
\end{split}
\end{equation}•
We now prove that this operator satisfies
\begin{equation}
\label{eq: property of q}
\begin{split}
q \left( \s = \omega^{\delta} = e^{\mathrm{i}\frac{2\pi}{n}\delta} \right)
=
\delta
\,,
\quad
\delta  = 0 , ... , n-1
\,.
\end{split}•
\end{equation}•

In order prove Eq.~(\ref{eq: property of q}), we start by showing that $q(\omega^{0}) = 0$:
\begin{equation}
\begin{split}
q(\omega^{0})
&\,=
\frac{n-1}{2}
+
\sum^{n-1}_{a=1}\,
\frac
{
1
}
{
\bar{\omega}^a - 1
}
\\
&\,=
\frac{n-1}{2}
+
\textrm{Re}
\Big[
\sum^{n-1}_{a=1}\,
\frac
{
\omega^a - 1
}
{
|\, \omega^a - 1 \,|^2
}
\Big]
\\
&\,=
\frac{n-1}{2}
+
\sum^{n-1}_{a=1}\,
\left( - \frac{1}{2} \right)
\\
&\,=
0
\,.
\end{split}•
\end{equation}•

Now it is simple to verify that the following relation holds:
\begin{equation}
\label{eq: induction in q}
\begin{split}
q(\bar{\omega}\s)
=
q(\s) + \sum^{n-1}_{a = 1}\,\s^{a}
\,.
\end{split}•
\end{equation}•
From the fact that 
$
\sum^{n-1}_{a = 1}\,\s^{a}
=
n-1
$
if
$
\s = \omega^{0} = 1
$
and
$
\sum^{n-1}_{a = 1}\,\s^{a}
=
-1
$
if
$
\s = \omega^{\delta}
$,
$
\delta \in \{1, ... , n-1 \} 
$,
then it is possible to use Eq.~(\ref{eq: induction in q}) to establish
Eq.~(\ref{eq: property of q}) by induction.

Thus combining Eq.~(\ref{eq: definition of q})
and Eq.~(\ref{eq: property of q})
yields
\begin{subequations}
\label{eq: Identity Zn operators}
\begin{equation}
\begin{split}
\exp
{
\left\{
\mathrm{i}\,\frac{2\,\pi p}{n}\,
\left[
\frac{n-1}{2}
+
\sum^{n-1}_{a=1}\,
\frac
{
\left(
\s^{\dagger}_{j}\,\s_{j'}
\right)^{a}
}
{
\bar{\omega}^{a} - 1
}
\right]
\right\}
}
=
\left(
\s^{\dagger}_{j}\,\s_{j'}
\right)^{p}
\,,
\end{split}•
\end{equation}•
\begin{equation}
\begin{split}
\exp
{
\left\{
-\mathrm{i}\,\frac{2\,\pi p}{n}\,
\left[
\frac{n-1}{2}
+
\sum^{n-1}_{a=1}\,
\frac
{
\left(
\s^{\dagger}_{j}\,\s_{j'}
\right)^{a}
}
{
\omega^{a} - 1
}
\right]
\right\}
}
=
\left(
\s^{\dagger}_{j}\,\s_{j'}
\right)^{p}
\,,
\end{split}•
\end{equation}•
\end{subequations}•

for $p \in \{ 0,...,n-1 \}$ (mod $n$).

\end{document}